\documentclass[prb,twocolumn,aps,showpacs,fixfloats]{revtex4}
\usepackage{graphicx}
\usepackage{subfigure}
\usepackage{float}
\usepackage{bm}
\usepackage{amsmath,amssymb}

\usepackage{latexsym}
\usepackage{color}
\usepackage{enumerate}
\usepackage{pdfpages}
\usepackage{tikz}
\usepackage{hyperref}

\begin{document}
\newcommand{\s}{\scriptscriptstyle}
\newcommand{\uu}{\uparrow \uparrow}
\newcommand{\ud}{\uparrow \downarrow}
\newcommand{\du}{\downarrow \uparrow}
\newcommand{\dd}{\downarrow \downarrow}
\newcommand{\ket}[1] { \left|{#1}\right> }
\newcommand{\bra}[1] { \left<{#1}\right| }
\newcommand{\bracket}[2] {\left< \left. {#1} \right| {#2} \right>}
\newcommand{\vc}[1] {\ensuremath {\bm {#1}}}
\newcommand{\tr}{\text{Tr}}
\newcommand{\Trans}{\ensuremath \Upsilon}
\newcommand{\Refl}{\ensuremath \mathcal{R}}


\title{High-temperature magneto-inter-chirality
oscillations in 2D systems with 
strong spin-orbit coupling}

\author
{M. E. Raikh}

\affiliation{ Department of Physics and
Astronomy, University of Utah, Salt Lake City, UT 84112}
\begin{abstract}
Conventional Shubnikov–de Haas oscillations of conductivity in
3D are washed out as the temperature exceeds the spacing between the Landau levels. This is due to smearing of the
Fermi distribution. In 2D, in the presence of two or more
size-quantization subbands, there is an additional type of
magneto-oscillations, usually referred to as magneto-inter-subband oscillations, which do not decay exponentially with temperature. The period of these oscillations is determined by the condition that the energy separation between the subbands contains an integer number of Landau levels. Under this condition, which does not contain the Fermi distribution,  the inter-subband scattering rate is maximal. Here we show that, {\em with only one subband}, high-temperature oscillations are
still possible. They develop when the electron spectrum is split due to the spin-orbit coupling. For these additional oscillations, the coupling enters {\em both}
the period and the decay rate.
\end{abstract}

\maketitle

\section{Introduction}

Experimental study of the quantum resistance oscillations is usually carried out by plotting the Fourier
transform of the field-dependent correction to the resistance versus the
inverse magnetic field, $B$. In two dimensions, the slope of the straight line obtained yields the information about the density of electrons occupying the corresponding size-quantization subband. If two or more subbands are occupied, the 
low-temperature Fourier transform exhibits two or more peaks. As the temperature, $T$, is elevated, the peaks quickly vanish. This is the result of smearing of the Fermi distribution. At elevated temperatures the Shubnikov–de Haas peaks give way to the additional peaks
corresponding to the sum and difference of the fundamental
frequencies. The peak corresponding to the difference 
does not decay with $T$ since 
the effects of smearing coming from both
subbands get cancelled as a result of the subtraction.
The underlying mechanism of the slow decaying peak 
corresponding to the difference of the fundamental frequencies
is the inter-subband scattering of electrons by the disorder.
This effect is confidently observed experimentally, see e. g.
Refs. [\onlinecite{Coleridge,Leadly,Bykov,Vitkalov,Vitkalov1,minkov}],
and understood theoretically in tiniest details. \cite{Polyanovsky,shahbazyan,Golub,Grigoriev,Raichev} 

A natural question arises: whether the high-temperature oscillations 
are possible if there is only one subband?
One option is the spin splitting of the single-subband spectrum. 
As argued in Ref. [\onlinecite{Bykov}], the situation mimicking the inter-subband scattering can be realized if the applied magnetic field is tilted with respect to the 2D plane. Then the
condition for magneto-oscillations, which plays the role of inter-subband scattering, is that the
Zeeman splitting in the parallel field contains an
integer number of Landau levels created by  the normal component of the net magnetic field.

A question addressed in the present paper:
what if magnetic field is {\em strictly normal} to the 2D plane? Below we demonstrate that strong enough spin-orbit coupling {\em alone} is sufficient for the disorder-induced magneto-oscillations. This is despite, in the
presence of spin-orbit coupling, the Landau levels are not even-spaced. A  "price" paid for inability to align the staircases of Landau levels from two spin-orbit branches is
that both the period of additional oscillations and their decay with temperature contain  the spin-orbit coupling strength.

\section{Hamiltonian and eigenfunctions}

At a zero magnetic field, the Hamiltonian of spin-orbit coupled 
2D gas has the form

\begin{equation}
\label{HAMILTONIAN}
{\hat H}= \frac{\hbar^2{\hat {\bm k}}^2}{2m} +
\alpha({\hat {\bm \sigma}} \times {\hat {\bm k}}    )
\cdot {\bm z}_0,
\end{equation}
where ${\bm z}_0$ is the normal to the 2D plane,
${\hat {\bm k}}$ is the momentum operator,  while $ {\hat {\bm \sigma}} $ is the vector of
Pauli matrices. Parameters $m$ and $\alpha$ stand for the effective mass and the spin-orbit constant, respectively.
The spectrum of the  Hamiltonian Eq. (\ref{HAMILTONIAN}) consists
of two branches
\begin{equation}
\label{branches}
E_{1,2}({\bm k})=\frac{\hbar^2k^2}{2m}\mp\alpha|{\bm k}|.
\end{equation}
The corresponding wave functions have the form
\begin{equation}
\label{wavefunctions}
\Psi_{{\bm k}}^{(1,2)}({\bm r})=e^{i{\bm k}\cdot{\bm r}}\chi_{{\bm k}}^{(1,2)},
\end{equation}
where the spinors $\chi_{{\bm k}}^{(1,2)}$ are defined as
\begin{equation}
\label{chi}
\chi_{{\bm k}}^{(1)}=\frac{1}{\sqrt{2}}
\begin{pmatrix}
e^{i\Phi_{{\bm k}}}\\
-1
\end{pmatrix}, \hspace{3mm}
\chi_{{\bm k}}^{(2)}=\frac{1}{\sqrt{2}}
\begin{pmatrix}
1\\
e^{-i\Phi_{{\bm k}}}
\end{pmatrix}
\nonumber
 \end{equation}
with $\Phi_{{\bm k}}$ being the azimuthal angle of the vector ${\bm k}$.
It is crucial that the lower branch of the spectrum 
Eq.~(\ref{branches}) has a minimum at
\begin{equation}
\label{k0}
k=k_0= \frac{m\alpha}{\hbar^2}.
\end{equation}
Near this minimum the spectrum can be simplified as
\begin{equation}
 \label{minimum}
 E_1({\bm k})=-\Delta +\frac{\hbar^2(k-k_0)^2}{2m},
\end{equation}
with the depth of the minimum 
$\Delta=\frac{m\alpha^2}{2\hbar^2}=\frac{\hbar^2k_0^2}{2m}$.



Switching on the magnetic field implies the replacement of ${\hat{\bm k}}$ by the kinetic momentum ${\hat{\bm k}}+\frac{e}{c}{\bm A}$, where ${\bm A}$ is the vector 
potential, which we choose in the form ${\bm A}={\bm y}_0Bx$.

In the absence of spin-orbit coupling, the eigenstates are the conventional Landau levels. The corresponding wave functions are $e^{ip_yy}\psi_{n}(x-p_yl^2)$, where $p_y$
is the conserving $y$-component of the momentum, $l=(\hbar c/eB)^{1/2}$ is the magnetic length, and $\psi_n$ is the harmonic oscillator wave function.
Since the operator ${\hat{\bm k}}$ couples only the neighboring Landau levels, the eigenstates in the the presence of both magnetic field and spin-orbit coupling 
 represent the following 
 vectors of the harmonic
oscillator wave functions\cite{Rashba}


\begin{equation}
\label{spinors}
\Psi\left(x,y\right)=e^{ip_yy}
\begin{pmatrix}
 i\lambda b_n^{-\lambda}\psi_{n-1}(x-p_yl^2)\\
 b_n^{\lambda}\psi_n(x-p_yl^2)
\end{pmatrix},
\end{equation}
where $\lambda$ denotes the chirality and
takes the values $\pm 1$.
Normalization 
coefficients $b_n$ are given by

\begin{equation}
\label{bn} 
b_n^{\lambda}=\Bigl(\frac{1}{2} +
\frac{\lambda}{4c_n}\Bigr)^{1/2}, 
\end{equation}
and the coefficients $c_n$ are
defined as 
\begin{equation}
\label{cn}
c_n=\Big(\gamma^2n+\frac{1}{4}\Big)^{1/2}.
\end{equation}
Spin-orbit constant, $\alpha$, enters into Eqs. (\ref{bn}) and (\ref{cn}) via the parameter $\gamma$
given by
\begin{equation}
\label{gamma}
\gamma=\Biggl[\frac{2m\alpha^2}{\hbar^2(\hbar\omega_c)}\Biggr]^{1/2}=
\Biggl(  \frac{4\Delta}{\hbar\omega_c}\Biggr)^{1/2},
\end{equation}
where $\omega_c=\Big(\frac{eB}{mc}\Big)^{1/2}$ is the 
cyclotron frequency.

Energy level positions corresponding to the eigenfunctions Eq. (\ref{spinors}) are equal to 
\begin{equation}
\label{levels}
E_{n}^{\lambda}
=\hbar\omega_c\big(n+\lambda c_n\big).
\end{equation}
Naturally, for vanishing spin-orbit coupling $\gamma \rightarrow 0$,
the values $c_n$ approach $\frac{1}{2}$, as it should be for
the textbook Landau levels.
It is seen from Eq. (\ref{levels}) 
that the spacings between the neighboring levels 
\begin{equation}
\label{spacings}
\frac{E_{n+1}^{\lambda}-E_n^{\lambda}}{\hbar\omega_c}
=1+\lambda\big(c_{n+1}-c_n   \big)
\end{equation}
depends on the chirality, $\lambda$. This fact is crucial
for the period and the decrement of the inter-chirality oscillations.

As it is seen from Eq. (\ref{gamma}), the dimensionless parameter $\gamma$
represents a square root of the number of the cyclotron quanta that fit into the minimum of the depth $\Delta=\frac{m\alpha^2}{2\hbar^2}$ in the zero-field
spectrum Eq. (\ref{minimum}), see Figure.


\section{Disorder-induced inter-chirality scattering, $E_F\gg \Delta$}
In the presence of disorder, $V({\bf r}) $, the discrete
energies, $E_n^{\lambda}$, get broadened. This broadening
is captured by the chirality-dependent self-energy, 
$\Sigma^{\lambda}(E)$. For a given chirality, the density of 
states is expressed via $\Sigma^{\lambda}(E)$ as
\begin{align}
\label{selfenergy}
&    g_{\lambda} (E)=\frac{1}{2\pi^2l^2}{\text  I}{\text m} \sum_n\frac{1}{E-E_n^{\lambda}-\Sigma^{\lambda}(E)}\nonumber\\
&=\frac{1}{2\pi^2l^2}\sum_n 
\frac{{\text  I}{\text m} \Sigma^{\lambda}  }{\big(E-E_n^{\lambda}\big)^2+
{\big(\text  I}{\text m} \Sigma^{\lambda}\big)^2}.
\end{align}
For a weak magnetic field $E\gg {\text  I}{\text m} \Sigma^{\lambda} \gg\hbar\omega_c$ the  summation can be
replaced by integration yielding
\begin{equation}
\label{summation}
  g_{\lambda}^{(0)} (E)=\Big(\frac{m}{2\pi\hbar^2}\Big)
  \frac{1}{1+\frac{\lambda\gamma}{2}
  \big(\frac{\hbar\omega_c}{E}    \big)}.
\end{equation}
Oscillating correction to the
density of states comes from the first harmonics of the 
Poisson expansion of Eq. (\ref{selfenergy})
\begin{align}
\label{Poisson}    
&g_{\lambda}^{(1)} (E)=2g_{\lambda}^{(0)} (E)\cos\Bigg[2\pi\frac{E}{\hbar\omega_c}    
+2\pi\gamma\lambda \Big( \frac{E}{\hbar\omega_c}     \Big)^{1/2}    \Bigg] \times\nonumber\\
&\exp\Bigg[-2\pi\frac{{\text  I}{\text m}\Sigma^{\lambda}(E)   }{\hbar\omega_c}     \Bigg].
\end{align}
The form of the argument of the cosine follows from the combination
$n+c_n$ in the expression Eq. (\ref{levels}) for $E_n^{\lambda}$
with $n=\frac{E}{\hbar\omega_c}$. Higher harmonics in $g_{\lambda}^{(1)}$ can be neglected 
under the condition ${\text  I}{\text m}\Sigma^{\lambda} \gg \hbar\omega_c$, which
allows to replace the summation over the Landau levels by integration.
Since the meaning of ${\text  I}{\text m}\Sigma^{\lambda}$ is the inverse
scattering time, this condition implies that the scattering time is much
shorter than the cyclotron period.

At temperature $T=0$ the inter-chirality magneto-oscillations originate from the
fact that inter-chirality scattering time is proportional
to the product of the two oscillating corrections
$ g_{1}^{(1)}(E) g_{-1}^{(1)}(E)  $ taken at the Fermi energy, 
$E=E_F$. From Eq. (\ref{Poisson}) we find

\begin{align}
\label{product}
& g_{1}^{(1)}(E_F) g_{-1}^{(1)}(E_F)=4 g_{1}^{(0)}(E_F) g_{-1}^{(0}(E_F)\nonumber\\    &\times\cos\Biggl[2\pi\frac{E_F}{\hbar\omega_c}+2\pi\gamma\Bigl(\frac{E_F}{\hbar\omega_c}   \Bigr)^{1/2}\Biggr]\nonumber\\
&\times \cos\Biggl[2\pi\frac{E_F}{\hbar\omega_c}-2\pi\gamma\Bigl(\frac{E_F}{\hbar\omega_c}   \Bigr)^{1/2}\Biggr]e^{-2W} 
\end{align}
where the Dingle factor, $e^{-2W}$, is defined as

\begin{equation}
e^{-2W}=\exp\Bigg[-2\pi\frac{{\text  I}{\text m}\Sigma^{1}(E) +
{\text  I}{\text m}\Sigma^{-1}(E)}{\hbar\omega_c}\Biggr].
 \end{equation}

We now convert the  product of cosines into a sum
\begin{align}
\label{sum}
&2\cos\Biggl[2\pi\frac{E_F}{\hbar\omega_c}+2\pi\gamma\Bigl(\frac{E_F}{\hbar\omega_c}   \Bigr)^{1/2}\Biggr]\nonumber\\
&\times\cos\Biggl[2\pi\frac{E_F}{\hbar\omega_c}-2\pi\gamma\Bigl(\frac{E_F}{\hbar\omega_c}   \Bigr)^{1/2}\Biggr]\nonumber\\
&=\cos\Biggl[4\pi\frac{E_F}{\hbar\omega_c}\Biggr]+\cos\Biggl[4\pi\gamma\Bigl(\frac{E_F}{\hbar\omega_c}   \Bigr)^{1/2}                 \Biggr].
\end{align}

 It follows from Eq. (\ref{sum}) that the inter-chirality scattering rate
 is the sum of two contributions both oscillating with magnetic field.
 Since the first term oscillates much faster than the second one, they
 decay with temperature at completely different rate.
 The way  to incorporate finite temperature into Eq. (\ref{sum})  is straightforward.
 One should replace $E_F$ by $E_F+Tz$ and integrate the result  together with the dimensionless
 Fermi function $(e^z+1)^{-1}$. For the rapidly oscillating term, corresponding to 
the sum,  this yields
\begin{equation}
\label{T}
\Bigl(g_{1}^{(1)} g_{-1}^{(1)}\Bigr)^{(+)}
=(g^{(0)})^2\cos\Biggl(\frac{4\pi E_F}{\hbar\omega_c}\Biggr)A\Biggl(\frac{T}{T_1}\Biggr)e^{-2W},
\end{equation}
where the conventional damping function $A(\chi)$ is defined as
\begin{equation}
\label{damping}
A(\chi)=\frac{\chi}{\sinh \chi},   
\end{equation}
where the argument, $\chi$, is given by
\begin{equation}
\label{CHI}
\chi=\frac{4\pi^2T}{\hbar\omega_c},
\end{equation}
i.e. $T_1=\frac{\hbar\omega_c  }{4\pi^2}$ which is two times smaller than $T_1$ for conventional 
Shubnikov-de Haas oscillations, while the period of the magneto-interchirality oscillations is
two times bigger than that for conventional oscillations.

Most interesting is the result of the temperature averaging of the 
second term in Eq. (\ref{sum}), corresponding to the difference. It reduces to the same function $A(\chi)$,
namely,
\begin{equation}
\label{TT}
\Bigl(g_{1}^{(1)} g_{-1}^{(1)}\Bigr)^{(-)}
=(g^{(0)})^2\cos\Biggl[4\pi\gamma\Bigl(\frac{E_F}{\hbar\omega_c}   \Bigr)^{1/2}\Biggr]A\Biggl(\frac{T}{T_2}\Biggr)e^{-2W},
\end{equation}
where $T_2$ is now given by
\begin{equation}
\label{T2}
 T_2=\frac{\Big(\hbar\omega_cE_F\Big)^{1/2}}
 {2\pi^2\gamma}.   
\end{equation}
Our most important observation is that Eq.
(\ref{TT}) describes high-temperature
oscillations, i.e. $T_2\gg T_1$. To demonstrate this, we rewrite the ratio 
$\frac{T_2}{T_1}$ in the form
\begin{align}
\label{ratio}
&\frac{T_2}{T_1}=2\frac{\left(\hbar\omega_cE_F   \right)^{1/2}}{\hbar\omega_c\gamma   }
=2\Bigl( \frac{E_F}{\hbar\omega_c}  \Bigr)^{1/2}\Biggl[\frac{\hbar^2(\hbar\omega_c)}{2m\alpha^2}\Biggr]^{1/2}
\nonumber\\
&=2\Biggl(\frac{\hbar^2E_F}{2m\alpha^2}    \Biggr)^{1/2}=2\Biggl( \frac{E_F}{\Delta}   \Biggr)^{1/2}.
\end{align}
We see that the condition $T_2\gg T_1$ is met when the Fermi level is much higher that 
the depth, $\Delta$,
of the spin-orbit minimum in the energy  spectrum at zero magnetic field.
It is also instructive to cast the "slow" cosine in Eq. (\ref{TT}) in the form
\begin{equation}
\label{argument}
  4\pi\gamma \Biggl(\frac{E_F}{\hbar\omega_c}   \Biggr)^{1/2} =
  4\pi \frac{\Bigl(E_F\frac{2m\alpha^2}{\hbar^2}     \Bigr)^{1/2}   }{\hbar\omega_c}=4\pi\frac{\varepsilon_0}{\hbar\omega_c}. 
\end{equation}
It follows from Eq. (\ref{argument}) that the same condition $E_F \gg \frac{2m\alpha^2}{\hbar^2}$, i.e. $\varepsilon_0 \ll E_F$,
which ensures that $T_2\gg T_1$ also  ensures that the "slow" cosine in Eq. (\ref{TT}) is indeed parametrically
slower than the cosine in Eq. (\ref{T}).

We assumed that the magnetic field is strictly normal to the 2D plane.
In addition to the Landau quantization, this field causes the Zeeman splitting
$\pm \Delta_Z$ of the electron spectrum.
It is straightforward to incorporate the Zeeman splitting into the above result for the period of slow
oscillations. Namely, in the presence of the Zeeman splitting, the oscillating
magnetoresistance $\cos4\pi\Bigl(\frac{\varepsilon_0}{\hbar\omega_c}\Bigr)$ transforms into
\begin{equation}
\label{transforms}
\cos\Biggl[4\pi\Bigl(\frac{\varepsilon_0}{\hbar\omega_c}\Bigr)\Bigl(1+\frac{\Delta_Z^2}
{\Delta_0^2}\Bigr)^{1/2}\Biggr].
\end{equation}
Characteristic Zeeman splitting $\Delta_0$ at which the frequency of magneto-oscillations
increases twice is given by
\begin{equation}
\label{twice}
\Delta_0\!=E_F^{1/2}\!\Biggl(\frac{2m\alpha^2}{\hbar^2}   \Biggr)^{1/4}\!\Bigl(\hbar\omega_c   \Bigr)^{1/4}\!=\!E_F\Biggl(\frac{\varepsilon_0}{E_F}  \Biggr)^{1/2}\Bigl(\frac{\hbar\omega_c}{E_F}  \Bigr)^{1/4}.
\end{equation}
We see that the last two brackets are small by virtue of the assumptions made. 
This leads us to the conclusion that, with Zeeman splitting, slow magneto-oscillations
survive, but become "faster". Obviously, the magnitude of the magneto-oscillations
falls off with $\Delta_Z$. This is because the larger is $\Delta_Z$ the smaller is the 
probability of the inter-chirality scattering.

The remaining task is to demonstrate that the term Eq.~(\ref{TT}) indeed emerges as a certain term in the expression for the conductivity. More
specifically, one has to trace how the product 
$\Bigl(g_{1}^{(1)} g_{-1}^{(1)}\Bigr)^{(-)}$ originates from the general
Kubo formula

\begin{equation}
\label{Kubo}
\sigma_{xx}(E)=\frac{\pi e^2 \hbar}{\Omega}
\langle   {\text  T}{\text r}\Bigl[{\hat v}_x\delta(E-{\hat H}-V({\bm r})){\hat v}_x\delta(E-{\hat H}-V({\bm r}))                      \Bigr]                    \rangle. 
\end{equation}
Here $\langle ...\rangle$ denotes the averaging over the disorder 
$V({\bm r})$, 
${\hat v}_x$ is the velocity operator,  and $\Omega$ is the 2D normalization
volume. Similarly to intersubband scattering\cite{Polyanovsky,shahbazyan,Golub,Grigoriev,Raichev}, 
evaluation of $\sigma_{xx}$ is possible neglecting
the localization effect when the self-consistent Born approximation (SCBA) applies. 
As shown in Ref. \onlinecite{shahbazyan}
this requires that  the disorder
is short-ranged, namely, that the correlation radius is much smaller than $l$.
In the framework of the SCBA, the $\delta$-functions in the right-hand side
of Eq. (\ref{Kubo}) can be averaged
{\em independently}. Within a factor $\frac{1}{\pi}$ each average $\delta$-function is expressed through $g_{\lambda}(E)$ and thus, via 
Eq. (\ref{selfenergy}), through the self-energy. This is how the product of
${\text  I}{\text m}\Sigma^{\lambda}$ with different $\lambda$ enters into
$\sigma_{xx}$.

The imaginary part
of the self-energy is determined by the level numbers,
$n$, for which $E_n^{\lambda}$ is close to $E$. Equating 
$E_n^{\lambda}$ to $E$, we find from Eq. (\ref{levels})

\begin{equation}
\label{nE}
n^{\lambda}(E)=\Biggl[\Biggl(\frac{E}{\hbar\omega_c}+\frac{\gamma^2}{4}   \Biggr)^{1/2} \pm 
\frac{\lambda\gamma}{2}      \Biggr]^2.
\end{equation}
We see that, at a given $E$, the  corresponding numbers of the $n$-values 
depend on the chirality, $\lambda$. As a result, the self-energy depends
on chirality. This fact is at the core of inter-chirality magneto-oscillations.

To conclude the Section, we relate $\Sigma^{\lambda} $ to the 
disorder strength. The quantitative characteristics of this strength
is the  width, $\Gamma$, defined as\cite{shahbazyan}

\begin{equation}
\label{Gamma}
\Gamma^2=\frac{1}{2\pi l^2}\int d{\bm r} \langle V(0)V({\bm r})\rangle.
\end{equation}

This width enters the system of SCBA equations for the self-energy

\begin{align}
\label{RPA}
&\frac{\Sigma^{\lambda}}{\Gamma^2}=\sum_n\frac{\kappa_{\lambda,\lambda}}
{E-E_n^{\lambda}-\Sigma^{\lambda}}+\sum_n\frac{\kappa_{\lambda,-\lambda}}
{E-E_n^{-\lambda}-\Sigma^{-\lambda}},
\end{align}
where $\kappa_{\lambda,\lambda}$ and $\kappa_{\lambda,-\lambda}$ are the squares
of matrix elements defined as
\begin{equation}
\label{elements}
\kappa_{\lambda,\lambda}={\Big\vert}\langle \lambda,n|V({\bf r})|\lambda,n\rangle
{\Big\vert}^2,
~~\kappa_{\lambda,-\lambda}={\Big\vert}\langle \lambda,n|V({\bf r})|-\lambda,n\rangle{\Big\vert}^2.
\end{equation}
Finally, we relate $\Gamma$ to the imaginary part of self-energy which enters the
Dingle factor:
${\text I}{\text m}\Sigma^{\lambda}(E)=\frac{\Gamma^2}{\hbar\omega_c}$.

\section{Intra-chirality scattering, $E_F<0$  and $|E_F|<\Delta$}

At negative energy, $E<0$, the states Eq. (\ref{levels}) exist only for chirality $\lambda=-1$,
however,  as illustrated  in the Figure, there are {\em two} values of $n$
corresponding to a given value of energy, namely

\begin{align}
\label{TWO}
&n_{-}(E)=\Biggl[\frac{\gamma}{2}+\Bigl( \frac{\gamma^2}{4}-\frac{|E|}{\hbar\omega_c}    \Bigr)^{1/2}   \Biggr]^2,\nonumber\\
&n_{+}(E)=\Biggl[\frac{\gamma}{2}-\Bigl( \frac{\gamma^2}{4}-\frac{|E|}{\hbar\omega_c}    \Bigr)^{1/2}   \Biggr]^2.
\end{align}

To find the oscillating component of the density of states,
we substitute Eq. (\ref{levels}) with $\lambda=-1$ into the 
expression Eq. (\ref{summation}) and apply the Poisson expansion.
For the first harmonics, this yields

\begin{equation}
\label{POISSON}
g_{-1}^{(1)}(E)\propto \int\limits_0^{\infty}\frac{ {\text  I}{\text m}\Sigma^{-1}(E)
\hspace{2mm}dx \cos2\pi x}{\biggl[E-\hbar \omega_c\Big(x-\gamma x^{1/2}\Big)\biggr]^2+\Big({\text  I}{\text m}\Sigma^{-1}(E)    \Big)^2}.
\end{equation}
To illuminate the structure of the integral Eq. (\ref{POISSON}), we make a 
substitution
\begin{equation}
\label{NEW}
    x=\gamma^2\Big(\frac{1}{2}+s   \Big)^2
\end{equation}
after which it assumes the form
\begin{equation}
\label{POISSON1}
g_{-1}^{(1)}(E)\propto \frac{ {\text  I}{\text m}\Sigma^{-1}(E)}{(\gamma \hbar\omega_c)^2}\int\limits_{-\frac{1}{2}}^{\infty}
\frac{ds (1+2s)\cos 2\pi\gamma^2\Big(\frac{1}{2}+s\Big)^2}
{\Biggl[\frac{E}{\gamma^2\hbar\omega_c}+\frac{1}{4}-s^2   \Biggr]^2
+\Big(\frac{{\text  I}{\text m}\Sigma^{-1}(E)}{\gamma^2\hbar\omega_c}    \Big)^2}.
\end{equation}
With the definition Eq. (\ref{gamma}) of the parameter $\gamma$, 
we realize that the square bracket in the denominator of Eq. (\ref{POISSON1}) turns to zero
at two values of the argument, namely
\begin{equation}
\label{PM}
s=s_{\pm}=\pm\Biggl(\frac{\Delta-|E|}{4\Delta}   \Biggr)^{1/2}.
\end{equation}
These values  contribute to the integral Eq. (\ref{POISSON1}) {\em independently} under the condition
\begin{equation}
\label{INDEPENDENTLY}
{\text  I}{\text m}\Sigma^{-1}
\ll \Bigl(\Delta-|E|\Bigr).    
\end{equation}
When this condition is met, the integral (\ref{POISSON1})
can be evaluated by expanding the integrand around $s=s_{+}$ 
and $s=s_{-}$. As a result, the oscillating component, 
$g_{-1}^{(1)}(E)$, represents a sum of two contributions
\begin{align}
\label{SUM}
&g_{-1}^{(1)}(E)\propto \frac{1}{\hbar \omega_c}
\Biggl[\cos\Biggl(\frac{8\pi\Delta}{\hbar\omega_c}\Biggr)
\Bigl(\frac{1}{2}+s_{+}\Bigr)^2\Biggr]\times \nonumber\\
&\exp\Big[-2\pi\frac{{\text  I}{\text m}\Sigma^{-1}}{\hbar\omega_c} \Big(\frac{1+2s_{+}}{2|s_{+}|}   \Big) \Big]+\nonumber\\
&\frac{1}{\hbar \omega_c}\Biggl[\cos\Biggl(\frac{8\pi\Delta}{\hbar\omega_c}\Biggr)
\Bigl(\frac{1}{2}+s_{-}\Bigr)^2\Biggr]\times\nonumber\\
&\exp\Big[ -2\pi\frac{{\text  I}{\text m}\Sigma^{-1}}{\hbar\omega_c}\Big(\frac{1+2s_{-}}{2|s_{-}|}   \Big) \Big].
\end{align}
Above we had established that in the case of inter-chirality scattering, slow-decaying oscillations
of the magneto-resistance originate from the {\em difference} of the
arguments of cosines in the  product  
$ g_{1}^{(1)}(E) g_{-1}^{(1)}(E)  $. On the contrary, for intra-chirality scattering, the slow-decaying oscillations originate from
the {\em sum} of the arguments of two cosines, which emerge in
the term $\propto g_{-1}^{(1)}(E)^2$ in the conductivity.
Indeed, the product of two cosines in Eq. (\ref{SUM})
contains a term
\begin{equation}
\label{TERM}   
\cos\Biggl[\Bigl(\frac{4\pi\Delta}{\hbar\omega_c}\Bigr)\Bigl(1 +\frac{\Delta-|E|}{\Delta} \Bigr)\Biggr].
\end{equation}
The latter expression leads us to the conclusion
that when the Fermi level is located close to the bottom
 of the spectrum with $\lambda=-1$,
i.e. for $\left(\Delta-|E_F|\right)\ll \Delta $,
the intra-subband correction to the conductivity 
oscillates as $\cos\Bigl(\frac{4\pi\Delta}{\hbar\omega_c}\Bigr)$ {\em as if} the concentration of electrons was $\frac{m\Delta}{\pi \hbar^2}$ (without spin-orbit coupling) rather than much smaller concentration $\frac{m(\Delta-E_F)}{\pi \hbar^2}$ (in the presence of a strong spin-orbit coupling). In other words, while the Fermi level is positioned near
the minimum of the spectrum, the conductivity 
{\em knows} about the {\em net depth of the minimum}, $\Delta$.

\section{Concluding remarks}
The origin of the conventional oscillations of magnetoconductivity 
$\delta \sigma \propto \cos\Bigl(2\pi \frac{ E_F}
{\hbar\omega_c}  \Bigr)$ is the fact that 
the Landau levels near the Fermi energy, 
$E_F$, are even-spaced. In this regard, the message of the present paper is that, spin-orbit coupling,
while lifting the equality of the spacings of the
Landau levels, 
leads to additional oscillations 
$\delta \sigma \propto \cos
\Bigl[ 8\pi\frac{ (\Delta E_F)^{1/2}}
{\hbar\omega_c}  \Bigr]$, where $\Delta$ is the
depth of the minimum in the energy spectrum, see Figure. These oscillations are much slower than the
conventional oscillations and decay, upon increasing temperature, much slower than
the conventional oscillations.

Recent experiments [\onlinecite{SmB6,E1,E3,E2,RRD,WTe,Young+Gate}]
had revived interest to the subject of magneto-oscillations in solids, see e.g. Refs.
[\onlinecite{1,2,P.A.Lee,Allocca,Allocca1,knolle}]. In short, the experiments [\onlinecite{SmB6,E1,E3,E2,RRD,WTe,Young+Gate}]
indicate that magneto-oscillations persist even when the Fermi level lies inside the
energy gap, so that the carriers participating in susceptibility and transport
are thermally excited. At the core of the proposed explanation is the observation
that, in materials exhibiting the non-conventional oscillations, the minimum of the
conduction band is located at finite momentum. A consequence of this fact is
a peculiar structure of the Landau levels, namely, as magnetic field increases,
one of the levels periodically approaches the band-edge, and, subsequently, narrows. This is
the situation very similar to that in Sect. IV.
In this regard, while in the  present manuscript, we considered the conventional 
"metallic" arrangement when  the Fermi level lies above the band minimum of the lower branch
of the spectrum Eq. (\ref{branches}) has a minimum at finite $k=k_0$. Then our prime
message is that the disorder-induced scattering between the levels with different $n$
boosts the oscillations.

\end{document}